\documentclass[a4paper]{jpconf}
\usepackage{graphicx} % Allows including images
\usepackage{amsmath}
\usepackage{slashed}
\usepackage{color}

\DeclareGraphicsExtensions{.jpg,.png,.pdf}

\begin{document}

\title{Excluded volume effects in the hybrid star EoS}

\author{David E. Alvarez-Castillo$^{1,2}$, Mark A. R. Kaltenborn$^{3}$, David Blaschke$^{1,3,4}$ } 
\address{Bogoliubov Laboratory for Theoretical Physics, JINR Dubna, 141980 Dubna, Russia}
\address{Universidad Aut\'{o}noma de San Luis Potos\'{i}, S.L.P. 78290 M\'{e}xico, Mexico}
\address{Instytut Fizyki Teoretycznej, Uniwersytet Wroclawski, 50-204 Wroclaw, Poland}
\address{National Research Nuclear University (MEPhI), 115409 Moscow, Russia}
%\ead{blaschke@ift.uni.wroc.pl}
%\ead{mkaltenborn@ift.uni.wroc.pl}
\ead{alvarez@theor.jinr.ru}

\begin{abstract}
In this contribution, we outline a new 2-phase description of the quark-nuclear matter hybrid equation of state that takes 
into account effects of phase space occupation (excluded volume) in both, the hadronic and the quark matter phases. 
For the nuclear matter phase, the reduction of the available volume at increasing density leads to a stiffening, while for 
the quark matter phase a reduction of the effective string tension in the confining density functional is obtained. The 
deconfinement phase transition in the resulting hybrid equation of state is sensitive to both excluded volume effects. 
As an application, we consider matter under compact star constraints of electric neutrality and $\beta$-equilibrium. We 
obtain mass-radius relations for hybrid stars that fulfill the $2M_\odot$ constraint and exhibit the high-mass twin 
phenomenon. Both features depend sensitively on the excluded volume.  
\end{abstract}

\section{Introduction}
There is a growing interest in the physics of compact stars, since their masses and radii are observables that can constrain the dense matter equation of state (EoS) at zero temperature and thus provide benchmarks, which are not accessible in terrestrial experiments. 
The question arises: is there quark matter in compact star interiors, and how would it manifest itself in observations? 
In this context, the recent conjecture about the existence of high-mass twin stars 
\cite{Alvarez-Castillo:2013cxa,Blaschke:2013ana}
is of special interest for the heavy-ion collision programmes searching for the critical point in the QCD phase diagram. 
If the pattern of a third family of compact stars, or just a quasi-horizontal branch in the population of high mass pulsars at $\sim2M_\odot$ in their mass-radius ($M-R$) diagram could be identified by means of precise mass and radius measurements, then this would imply the existence of a strong $1^{\rm st}$ order phase transition (PT) in compact star matter at $T=0$, which in turn proves the existence of at least one critical endpoint in the QCD phase diagram.  

As it has been shown in \cite{Benic:2014jia}, there are three necessary conditions on the hybrid star EoS in order to obtain high-mass twin configurations in the $M-R$ diagram:
(i) a strong stiffening of nuclear matter at supersaturation densities, 
(ii) a density dependent stiffening of quark matter 
and (iii) a sufficiently soft behaviour of quark matter at the deconfinement PT in order to provide the necessary large jump in energy density in order to render the hybrid star branch disconnected from the hadronic one. 
For the EoS criteria on hybrid star branches being connected or disconnected to hadronic ones, see also 
\cite{Alford:2013aca}.

%Modern approaches to the deconfinement transition should be based on chiral quark models

In this contribution, we present an effective relativistic density functional approach to quark-nuclear matter at zero temperature that implements a mechanism of quark confinement based on the Cornell-type confining potential. 
To this end, we revive the ideas of the string-flip model (SFM) \cite{Horowitz:1985tx,Ropke:1986qs} giving them a relativistic form as in \cite{Khvorostukin:2006aw}. 
In a simplified form such a model has recently already been applied to study massive hybrid 
stars \cite{Li:2015ida}. 
The main new ingredient of the present work is a heuristic ansatz for the reduction and vanishing 
of the effective string tension in dense matter, which is assumed to scale with the available volume fraction. This will provide condition (ii) while (i) is realized by an excluded volume prescription \cite{Typel:2015} applied to the DD2 EoS \cite{Typel:2005ba,Typel:2009sy} 
and (iii) from multiquark interactions \cite{Benic:2014iaa}. 
We will provide the two-flavor quark-nuclear matter EoS under neutron star conditions of charge neutrality and $\beta$-equilibrium and demonstrate the possibility to obtain high-mass twin 
stars with a third family branch that covers masses of $1.8$~--~$2.3~M_\odot$ and radii of $11$~--~$14$ km that are well separated 
from the radii of the hadronic star branch of $15.0$~--~$15.5$ km in the same mass range.

\section{Description of the quark matter model}

The quark phase of matter in compact stars is considered 
to be a combination of up, down quarks in $\beta$-equilibrium with electrons. 
We neglect the contribution from strange quarks since, due to the large threshold chemical potential 
of $\sim 1500$MeV, they are insignificant. 
Here, we consider quark confinement to be achieved through a density dependence of the quark masses 
$m_{\textrm{I}}=\Sigma_{\textrm{s}}$, generated by the scalar quark selfenergy $\Sigma_{\textrm{s}}$
that diverges for densities approaching zero.
We notice that in the equation of motion for hadrons the diverging quark selfenergies are exactly compensated by those of the confining interaction \cite{Glozman:2008fk}.

As usual for fermions, there is also a vector selfenergy $\Sigma_{\textrm{v}}$  that results in an energy correction contributing to the Hamiltonian functional,
\begin{equation}
	H_{QCD} = H_{\textrm{k}} + \sum_{q=u,d} m_{q_{0}}\bar{q}q + H_{\textrm{I}} 
	\equiv \sum_{q=u,d}\bar{q}\left(\gamma_0~\vec{\gamma}\cdot\vec{p}+m_{q}-\gamma_0\omega_0\right) q~,
\end{equation}
where $m_{q_{0}}$ ($q=u,d$) are the bare quark masses, which we neglect here in considering the chiral limit. $H_{\textrm{k}}$ is the kinetic term, and $H_{\textrm{I}}$ is the interaction term. 

The effective mass, $m_{q}$, and the vector meanfield, $\omega_0$, embody all the medium effects of interacting quark matter. 
At $T=0$, the density functional for the effective mass is taken from the string flip model 
(see also \cite{Li:2015ida}),
\begin{equation}
	m_{q} %\equiv m_{q_{0}} + m_{\textrm{I}} 
	= m_{q_{0}} + \Sigma_\textrm{s} = m_{q_{0}} +  \underset{\textrm{confinement}}{\underbrace{D(n_\textrm{B}) n_{\textrm{B}}^{-\frac{1}{3}}}} + \underset{\textrm{Coulomb or OGE}}{\underbrace{C n_{\textrm{B}}^{\frac{1}{3}}}},
\end{equation}
where OGE stands for one gluon exchange, and $\Sigma_{\textrm{s}}$
%$m_{\textrm{I}}$ is the interacting mass, 
is parameterized as a function of the baryon number density $n_{\textrm{B}}$. 
The confinement interaction dominates at lower densities, while the perturbative OGE interaction becomes more important at higher densities and is neglected for now. 
The main new element of the present study is $D(n_B)$, the effective in-medium string tension resulting from multiplying the vacuum string tension $\sigma\sim D_0$ between quarks with the available volume fraction $\Phi(n_\textrm{B})$,
\begin{eqnarray}
	D(n_\textrm{B}) &=& \underset{\textrm{effective string tension}}{\underbrace{{D}_0 ~ \Phi(n_{\textrm{B}})}} ~,\\
	\Phi(n_{\textrm{B}}) &=& \exp\left\{ -\frac{\alpha}{2}(n_{\textrm{B}}-n_0)\left(n_{\textrm{B}}-n_0+|n_{\textrm{B}} -n_0|\right)\right\} ~,\qquad\textrm{Gaussian}\\
	\Phi(n_{\textrm{B}}) &=& \left(1+v_{\textrm{ex}}n_{\textrm{B}}\right)^{-1}~, \qquad \qquad\qquad\qquad\qquad\qquad\textrm{inverse linear} 
	%\Phi(n_{B}) &= \left\{\underset{0, n_{B} > \frac{1}{V_{\textrm{ex}}}}{1-V_{\textrm{ex}} n_{B}, n_{B} < \frac{1}{V_{ex}}}\right.
\end{eqnarray}
where $v_{\textrm{ex}}$ is the excluded volume parameter and  $\alpha = v_{\textrm{ex}}^2$. 
The effect of the Gaussian versus the inverse linear scheme on the EoS are compared in Fig.~1. 
From~\cite{Benic:2014iaa}, we have
\begin{equation}
	\omega_{0}(n_{\textrm{B}}) = \frac{a}{\Lambda^{2}}n_{\textrm{B}} + \frac{b}{\Lambda^{8}} n_{\textrm{B}}^{3}~,
\end{equation}
where $\Lambda=200$ MeV has been introduced to allow $a$ and $b$ to be dimensionless quantities. 
The quark quasiparticle dispersion relation in quark matter is then
\begin{eqnarray}
	E_{q}(p) &=& \sqrt{p^{2} + m_{q}^{2}(n_{\textrm{B}})} + \omega_{0}(n_{\textrm{B}})~.
\end{eqnarray}
The EoS follows from the energy density  $\varepsilon = \sum_{q=u,d} \varepsilon_{q}$, with
the partial energy densities
\begin{eqnarray}
	\varepsilon_{q} &=& \frac{3}{\pi^{2}}\int_{0}^{p_{q}}dp ~p^{2} E_{q}(p) = \underset{\varepsilon_{\textrm{FG},q}}{\underbrace{\frac{3}{\pi^{2}} \int_{0}^{p_{q}}\sqrt{p^{2} + m_{q}^{2}(n_{\textrm{B}})} ~p^{2} dp}} + \underset{\Delta\varepsilon_{q}}{\underbrace{n_{q} \omega_{0}(n_{\textrm{B}})}}~,
\end{eqnarray}
where $p_{q}$ is the quark Fermi momentum. 
Since the quark masses $m_{q}$ are density dependent, the quark chemical potentials $\mu_{q}$ have an additional term, $\Delta \mu_{q}$ with respect to the free Fermi gas
\begin{eqnarray}	
	\mu_{q} &=& \frac{d\varepsilon}{dn_{q}} = \frac{\partial \varepsilon_{\textrm{FG},q}}{\partial p_{q}} \frac{\partial p_{q}}{\partial n_{q}} + \sum_{j}\left(\frac{\partial \varepsilon}{\partial m_{j}}\frac{\partial m_{j}}{\partial n_{q}} + \frac{\partial\Delta\varepsilon_{q}}{\partial n_{q}}\right)= \sqrt{p_{q}^{2} + m_{q}^{2}(n_{\textrm{B}})} + \Delta\mu_{q}~,\\
%\end{eqnarray}
%The linear relationship between $n_{j}$ and $n_{B}$ allows us to supplant one with the other in the above interaction chemical potential.
%From here,
%\begin{equation}
%\begin{split}
	\Delta\mu_{q} &=& 
	\sum_{i=u,d}\frac{3}{\pi^{2}}\int_{0}^{p_{i}}~p^{2}dp\frac{m_{i}}{\sqrt{p^{2} + m_{i}^{2}}} 
	%\underset{K_{m,q}(n_{\textrm{B}})}{\underbrace{
	\frac{\partial m_{i}}{\partial n_{q}}
	%}} 
	+ \omega_{0}(n_{\textrm{B}})+n_{\textrm{B}}\frac{\partial\omega_{0}(n_{\textrm{B}})}{\partial n_{\textrm{B}}}.
%\end{split}
%\end{equation}
\end{eqnarray}
The relevant chemical potentials satisfy the $\beta-$equilibrium condition  $\mu_{d}=\mu_{u}+\mu_{e}$ , where we assume that neutrinos escape without interacting.
The number densities are then constrained by the charge neutrality condition in compact stars 
$(2/3)n_u-(1/3)n_d-n_e=0$. 
With these values we are able to calculate total pressure as the thermodynamic potential of the system,
%and complete our EoS,
\begin{eqnarray}
	P_{\textrm{tot}} =-\sum_{i=u,d,e}\varepsilon_{i} +\sum_{i=u,d,e}\mu_{i}n_{i}~.
\end{eqnarray}

\begin{figure}[!htb] % p over n %
%\begin[L]{minipage}{0.5\textwidth}
\includegraphics[scale=0.38]{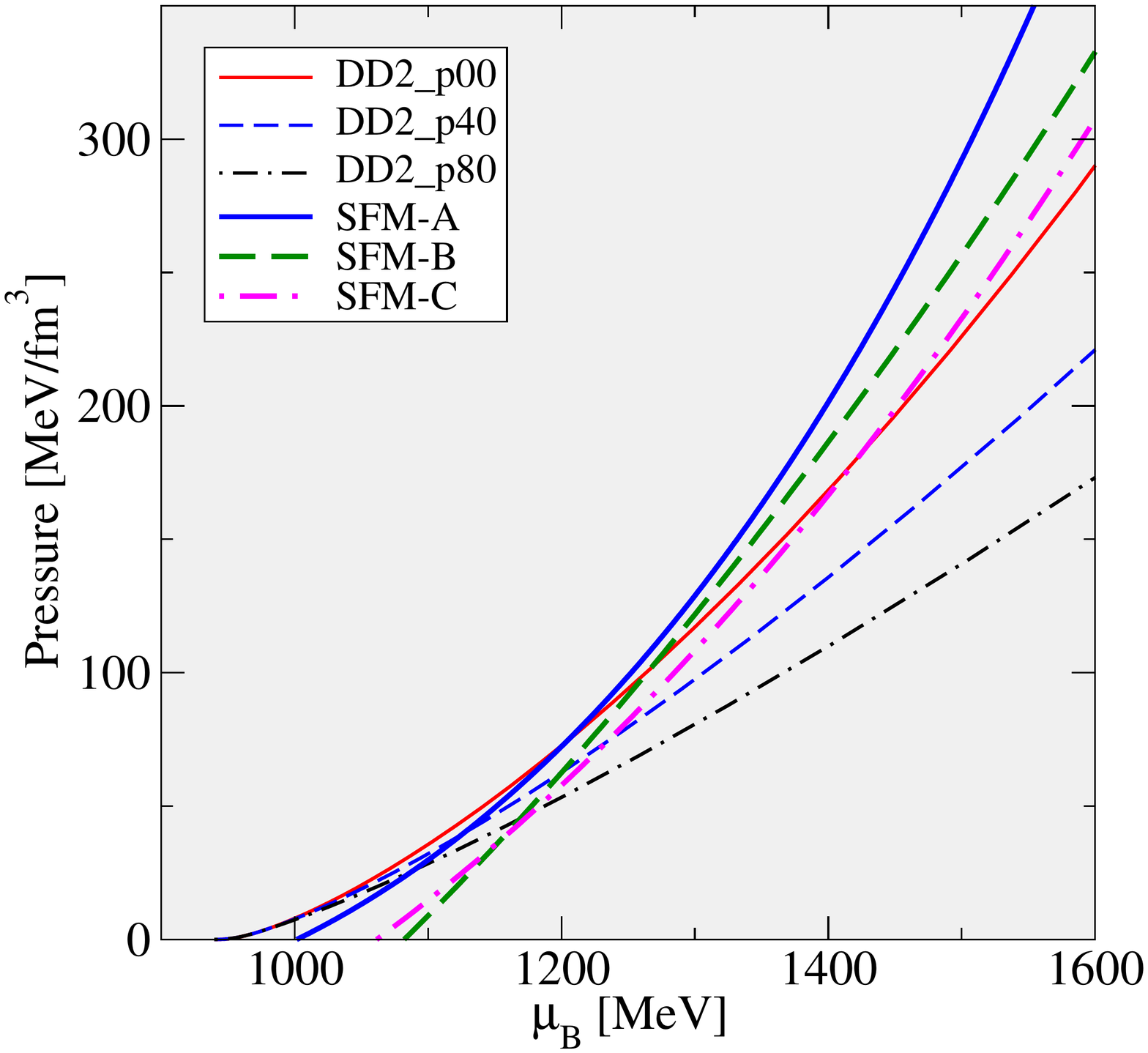} 
%\end{minipage}\hfill
%\begin[R]{minipage}{0.5\textwidth}
\hspace{-2.8cm}
\includegraphics[scale=0.38]{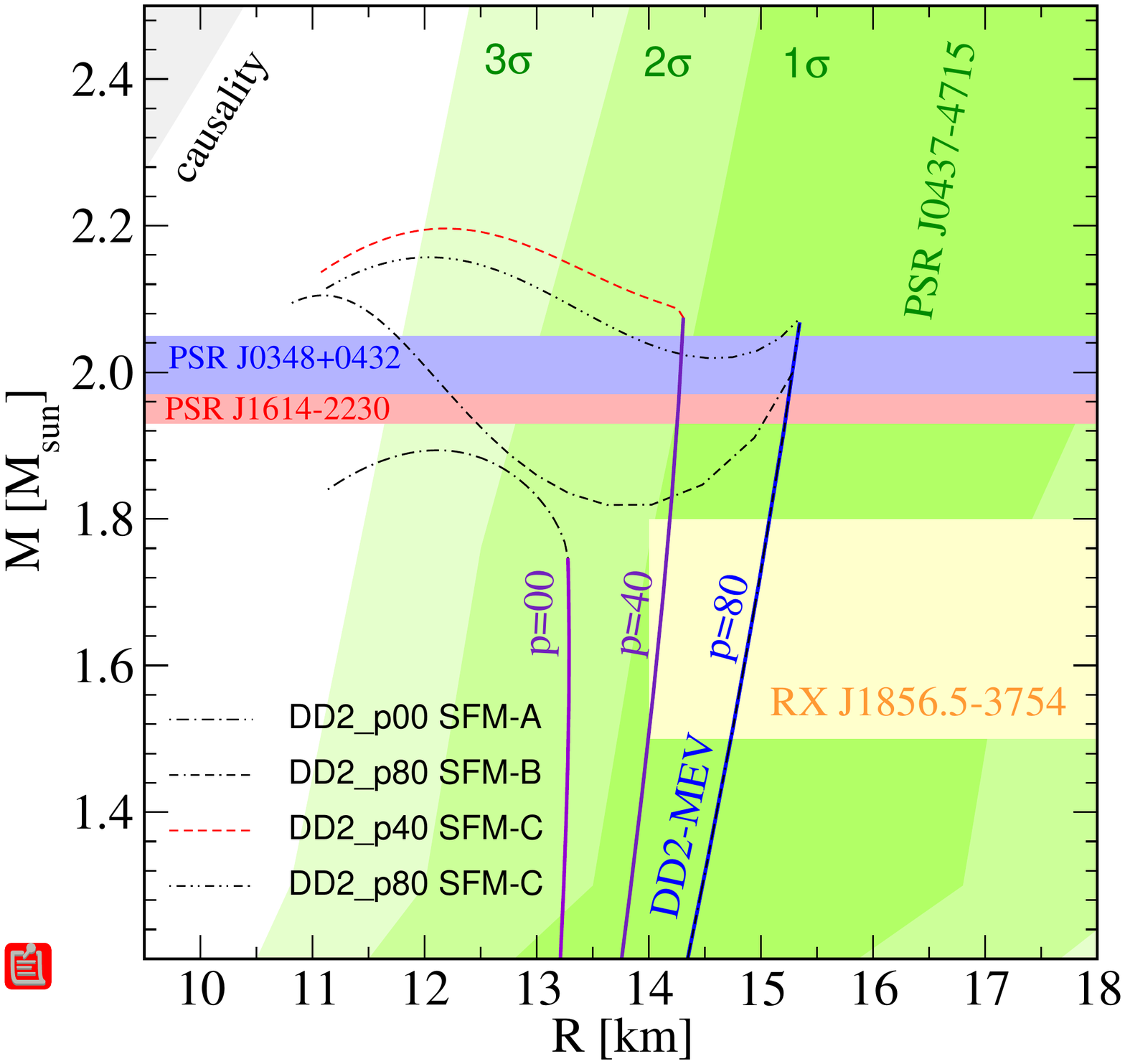} 
%\end{minipage}
\caption{Left: EoS with deconfinement transition between DD2-MEV (thin lines) and SQM (bold lines) 
for different excluded volume parameters, ($v_{\textrm{ex}}$[fm$^3$], $D_0$[MeV], a, b): SFM-A ($0.0$, $170$, $0.0$, $0.0$), SFM-B Gaussian $\Phi(n_\textrm{B})$ ($2.0$, $190$, $0.29$, $0.01$) and 
SFM-C inverse linear $\Phi(n_\textrm{B})$  ($2.0$, $220$, $0.25$, $0.04$). 
The label "p80" stands for positive $v_{\rm ex}=8.0$ fm$^3$, "p40" for $v_{\rm ex}=4.0$ fm$^3$, and 
"p00" for $v_{\rm ex}=0$.
Right: Mass-Radius sequences for compact stars with the EoS of the left panel. 
The hadronic part of the sequences is shown by the solid lines, the broken lines correspond to hybrid stars whereby the branches with positive slope are formed by unstable configurations.}
\label{fig:eos-mr}
\end{figure}

\section{Results and Conclusions}

Our results for the compact star EoS are shown in the left panel of Fig.~1 and demonstrate that increasing the excluded volume parameter results in a stiffening of the nuclear matter and a softening of the quark matter branches so that at the intersection of both branches in the $P-\mu_B$ diagram the change in pressure slope increases. 
This corresponds to a larger jump in the energy density at the transition and thus
to enforcing the instability induced by it for compact star configurations. 
The masses and radii for compact stars  are obtained from solving the Tolman-Oppenheimer-Volkoff equations with the corresponding hybrid EoS as input. 
Results are shown in the right panel of Fig.~1 and illustrate the above statement that the increase of the excluded volume parameter $v_{\rm ex}$ leads to the occurrence and prominence of the high-mass twin star effect. 
For  $v_{\rm ex}=0$ the $2M_\odot$ mass constraint \cite{Demorest:2010bx,Antoniadis:2013pzd}
is not fulfilled.

We have presented an effective relativistic density functional approach to quark-nuclear
matter at zero temperature that implements a mechanism of quark confinement based on the Cornell-type 
confining potential, thus reviving the ideas of the string-flip model in a relativistic fashion.
We have demonstrated that the main new ingredient of the present work, the effective reduction of the 
string tension in dense matter by the available volume fraction $\Phi(n_B)$ results in a softening of quark matter at  intermediate densities in the region of deconfinement transition. 
This provided a large jump in energy density at the transition as a necessary condition for obtaining high-mass twin star configurations and an extended branch of compact, massive hybrid stars with
masses of $1.8$~--~$2.3~M_\odot$ and radii of $11$~--~$14$ km that are well separated from the hadronic star branch in the same mass range.
It is a goal of future observational programmes to make sufficiently accurate neutron star radius measurements feasible.

\section*{Acknowledgements}
The work of D.E.A.-C. and D.B. was supported by Narodowe Centrum Nauki (NCN) under contract number 
UMO-2014/13/B/ST9/02621 while M.A.R.K. received support from NCN under contract number
UMO-2011/02/A/ST2/00306. D.E.A.-C. acknowledges the Heisenberg-Landau and Bogoliubov-Infeld programmes for supporting in part his collaboration visits at GSI Darmstadt (Germany) and at University of Wroclaw (Poland), respectively. 

\section*{References}
%\bibliography{proceedings}

\end{document}